\documentclass[journal]{IEEEtran}
\usepackage{graphicx}
\usepackage{eurosym}
\usepackage{microtype}
\usepackage{amssymb}
\usepackage{yfonts}
\usepackage{url}
\usepackage{multirow}
\usepackage{xcolor}
\usepackage{blindtext}
\usepackage{eucal}
\usepackage{enumitem}
\usepackage{amsmath}
\usepackage{algorithmic}
\usepackage{nomencl}
\usepackage{etoolbox}
\usepackage{eurosym}
\usepackage{amsfonts}
\usepackage{array}
\usepackage{comment} 
\makenomenclature
\usepackage{booktabs}
\usepackage{cite}
\usepackage{longtable}
\usepackage{threeparttable}
\usepackage{amsfonts} 
\usepackage{times}
\usepackage{nomencl}
\usepackage{subfigure}

\usepackage[caption=false]{subfig}

\makenomenclature

\makeatletter
\newcommand{\thickhline}{%
    \noalign {\ifnum 0=`}\fi \hrule height 0.75pt
    \futurelet \reserved@a \@xhline
}
\newcolumntype{"}{@{\hskip\tabcolsep\vrule width 0.75pt\hskip\tabcolsep}}
\makeatother

\makeatletter
\newcommand*{\rom}[1]{\expandafter\@slowromancap\romannumeral #1@}
\makeatother
\usepackage{supertabular}
\usepackage{array}
\usepackage{setspace}
\newenvironment{myenv}[1]
{\begin{spacing}{#1}}
    {\end{spacing}}
\usepackage{anysize}
\marginsize{10mm}{10mm}{10mm}{10mm}

\begin{document}

\title{Enhanced Transmission and Distribution Network Coordination to Host More Electric Vehicles and PV}
\author{Abbas~Rabiee, Andrew Keane and Alireza~Soroudi,
\thanks{
Abbas Rabiee, Andrew Keane, Alireza Soroudi (\{Abbas.rabiee, andrew Keane, alireza.soroudi\}@ucd.ie) are with the School of Electrical and Electronic Engineering, University College Dublin, Ireland. 
}
}

\markboth{}%
{Shell \MakeLowercase{\textit{et al.}}: Bare Demo of IEEEtran.cls for Journals}
\maketitle

\begin{abstract}
Distributed energy resources (DERs) installed at electric distribution networks create different opportunities and challenges for the distribution system operator (DSO). By increasing the penetration level of DERs, the impacts of these technologies are also sensed by the transmission system operator (TSO). The focus of this paper is on investigating and managing the impacts of the Electric Vehicle (EV) charging and solar Photo-Voltaic (PV) generation considering TSO-DSO interaction constraints. These constraints include the driving point impedance that reflects the impact of the DER-rich distribution system on the upstream HV network and voltage profile control through the on-load tap changing (OLTC) transformer's operation.
 The aim is to optimally utilize the existing assets and flexibilities to have more EVs and PVs at the distribution level without violating constraints at both transmission and distribution levels. 
The proposed model could readily be used by policymakers to design and implement the proper time-of-use tariffs for improved handling of the increasing EVs and PVs penetrations in emerging distribution networks. The proposed method is implemented on the IEEE 69-bus standard MV distribution network to demonstrate the applicability. 
\end{abstract}
\vspace{-2mm}
\section*{NOMENCLATURE}
\begin{myenv}{1}
    \begin{supertabular}{>{\arraybackslash}p{1.8cm} >{\arraybackslash}p{6.6cm} }
        \textbf{Abbreviations:} & \\
        $\textit{ATC}$& Available transfer capacity.\\
        $\textit{DSO}$&Distribution system operator.\\
        $\textit{TSO}$&Transmission system operator.\\
        $\textit{DER}$&Distributed energy resource.\\
        $\textit{OLTC}$&On-load tap changer.\\
        $\textit{HC}$&Hosting capacity.\\
        $\textit{EV}$&Electric vehicle.\\
        $\textit{EVC}$&EV charger.\\
        $\textit{PV}$&Photo-Voltaic.\\
        \textbf{Sets/Indices:} & \\
        $e$&Index for EVC type (residential or public).\\
        $b,k$&Index for electrical network buses.\\
        $t$&Index for operation intervals.\\
        $s$&\textcolor{black}{Index for uncertainty scenarios.}\\
        $\Omega_B$ & Set of network buses.\\
        $\Omega_{B_S}$&Set of MV network's boundary buses connected to the upstream network via OLTCs.\\
        $\Omega_{B_b}$&Set of network buses connected to bus $b$.\\
        $\Omega_{EV}$&Set of EVC types (residential or public).\\
        $\Omega_{B_{EV}}$&Set of network buses with EVCs.\\
        $\Omega_{B_{PV}}$&Set of network buses with PV units.\\
        $\Omega_{T}$&Set of time intervals.\\
        $\Omega_{S}$&\textcolor{black}{Set of all scenarios.}\\
        \textbf{Variables:}&\\
        $\Lambda$&Lower bound of the network's HC for EVs.\\
        $(PG/QG)_{b,t,s}$&Active/reactive power generation in node $b$ at time $t$ and scenario $s$, (pu).\\
        $(PD/QD)_{b,t,s}$&Active/reactive demand in bus $b$ at time  $t$ and scenario $s$, (pu).\\
        $(PV/QV)_{b,t,s}$&Active/reactive power generation by the PVs connected to bus $b$ at time $t$ and scenario $s$, (pu).\\
        $PE^{e}_{b,t}$&Active/reactive power demand of the EV type $e$ connected to bus $b$ at time $t$, (pu).\\
        $QE^{e}_{b,t,s}$&Reactive power exchange of the EV type $e$ connected to bus $b$ at time $t$ and scenario $s$, (pu).\\
        $(P/Q)_{bk,t,s}$&Active/reactive power flowing through $bk$-th line at time $t$ and scenario $s$, (pu).\\
        $I_{bk,t,s}$&Current flowing through $bk$-th line at time $t$ and scenario $s$, (pu).\\
        $\theta_{bk,t,s}$& Angle difference between the voltages of buses $b$ and $k$ at time $t$ and scenario $s$, (rad).\\
        $V_{b,t,s}$&Voltage magnitude of bus $b$ at time $t$ and scenario $s$, (pu).\\
        $\varepsilon^{e}_{t}$&Modified demand profile of EV type $e$ at $t$.\\
        $T_{t,s}$&Tap ratio of the OLTC connecting HV and MV networks at time $t$ and scenario $s$.\\
        $Z_{in,t,s}$&Impedance seen at the HV side of the OLTC connecting HV and MV networks, at time $t$ and scenario $s$.\\
        \textbf{Parameters:}&\\
        $\alpha$&Impedance scaling factor.\\
        $(\hat{PD}/\hat{QD})_{b,t,s}$&Nominal active/reactive demand in bus $b$ at time $t$ and scenario $s$, (pu).\\    
        $\hat{V}_{b,t,s}$&Nominal voltage magnitude of bus $b$ at time $t$ and scenario $s$, (pu).\\
        $(\overline{PG}/\underline{PG})_{b,t}$& Upper/lower limit of active power generation in bus $b$ and time $t$, (pu).\\
        $(\overline{QG}/\underline{QG})_{b,t}$& Upper/lower limit of reactive power generation in bus $b$ and time $t$, (pu).\\
        $\overline{I}_{bk}$&Upper limit of the current flowing through $bk$-th line, (pu).\\
        $\Gamma_{t,s}$&Available percentage of PV at time $t$ and scenario $s$.\\
        $CPV_b$&Nominal capacity of PV connected to bus $b$.\\
        $CEV_b$&Nominal capacity of EVC connected to bus $b$.\\
        $KP_{Z/I/P,b}$&Constant impedance (Z)/current (I)/power (P) coefficient of active power demand at bus $b$.\\
        $KQ_{Z/I/P,b}$&Constant impedance (Z)/current (I)/power (P) coefficient of reactive power demand at bus $b$.\\
        $(g/b)_{bk}$&conductance/susceptance of the line connecting buses $b$ and $k$.\\
        $\kappa_{PV/EV}^{+/-}$&Upper/lower coefficient limiting the reactive power exchange of the PVs/EVs.\\
        $V^{spc}_{b,t}$&Voltage set-point of OLTC's downstream bus.\\
        $(\overline{T}/\underline{T})$&Upper/lower limit of the OLTC's tap ratio.\\
        $Z_{min}$&Lower limit of the impedance seen via the OLTC's HV side.\\
    \end{supertabular}
\end{myenv}    

\vspace{-3mm}

\section{Introduction}\label{sec1}
\subsection{Background and Motivations}
\textcolor{black}{
In the transition from the traditional to modern electric power systems, distributed energy resources (DERs) are the key equipment employed to produce or store electricity closer to the end-users of electric power. The use of DERs may offer numerous benefits, including avoided generation/transmission capacity expansion costs, less need for backup power, and reduced emissions; but, it may also pose operational and economic challenges to electric utilities and their customers.}
\textcolor{black}{Additionally, due to the global warming issue as the most challenging environmental problem in the next decades, all future energy scenarios worldwide \cite{iea2020, mintzer2003us, ukfes, EirGrid}, emphasised the vital role of electrified transport in enabling the next phases of the energy transition towards high penetrations of sustainable energy sources. 
Additionally, among various types of DERs, the PVs and EVs gained much attention in recent years that can be considered as an integral part of the emerging smart electric distribution systems.
The charging and discharging power of EVs that can be managed flexibly and efficiently, can provide various regulatory benefits to distribution systems \cite{6919255}.
}

In normal conditions, the EV owners tend to charge their vehicles when required and PV units will generate when solar radiation is high enough \cite{DANESHVAR2020119267}. In high penetration levels of PVs and EVs, simultaneous EV charging or PV production will create technical challenges for both transmission and distribution networks. Some of them are listed here:\\
\begin{itemize}
\item Distribution level: Voltage profile, power curtailment of PVs, power losses and overloading the feeders \cite{9064904}.
\item Transmission level: Equivalent Impedance seen by TSO from the distribution network, voltage stability and transformer loading 
\end{itemize}
\subsection{Literature survey}
The coordination between TSO and DSO is necessary for better operation of both transmission and distribution systems \cite{7491385}. In \cite{gonzalez2020smart} the concept of smart power cell is developed to control the distribution system's active and reactive power exchange with the transmission network following set points provided by a superimposed control system.
The distribution network's flexibility ranges (seen by the TSO) in terms of active and reactive power provision capability at its boundary nodes with transmission system, is derived in \cite{kalantar2019characterizing}. \textcolor{black}{Besides, \cite{8291006} further developed active/reactive capability of active distribution network by introducing the flexibility cost maps, which allow the visualization of the impact of DERs flexibility at the interface of TSO-DSO}.
The coordinated voltage control of TSO and DSO in presence of DERs is another issue that gained the attention of many researchers \cite{sun2019review}. As an example, in \cite{9145713} a centralized OPF-based control scheme that
optimizes the real-time operation of active distribution network, while also considering the provision of voltage support as an ancillary service by the DERs. Also, \cite{8066335} analyzes the impact of DERs on the static voltage stability of an integrated transmission-distribution system via a continuation power flow analysis. \textcolor{black}{In \cite{9165851} a decentralized method is provided for EV charging to flatten the peak load caused by EV charging. Besides, \cite{7164328} investigates the capability of using EVs for distribution system voltage management, in coordination with OLTCs to mitigate the voltage problems caused by PVs. In \cite{7024906} a two-step framework is developed for scheduling EVs, aims to limit the burden on distribution and transmission assets as well as to ensure the charging of all EVs. In the first step, the number of vehicles to be charged during each hour is optimized, while in the second step the maximum number of vehicles that can be charged based on operating conditions during the next hour to ensure distribution reliability requirements, is determined.} 

The electrification of transportation is one of the main steps for energy decarbonisation in the future which poses more pressure on distribution networks. A comprehensive literature survey is presented by \cite{shaukat2018survey}, where the main challenges brought by transportation electrification for power systems are identified as power quality, economy, reliability \cite{shafiq2020reliability}, control, and grid load capacity. Finding the optimal EV charging pattern has been investigated in different works for different objectives such as:
\begin{itemize}
    \item Mitigating voltage unbalance \cite{8981892}.
    \item Maximizing renewable generation utilization \cite{yoon2017economic}.
    \item Reducing the volatility of renewable energy sources \cite{GONG2020665}.
    \item \textcolor{black}{Resilience enhancement of distribution networks \cite{8988201, 8386670}.}
    \item Minimization of operation costs and optimization of the power demand curve \cite{lu2018multi, li2020economic}. 
    \item Three aims, namely: demand-side management, imported power from the transmission level and reducing the emissions \cite{humfrey2019dynamic}. 
    \item Consumption tariffs to displace the EV charging to periods with lower network utility and major PV injections \cite{lagarto2017optimisation}. 
    \item EV demand modification considering the network constraints such as transformer and line limitations  \cite{de2014optimal}.
    \item Risk and uncertainty handling \textcolor{black}{\cite{aliasghari2020risk, 9069461, gengcoordinated}}.
\end{itemize}

As it is observed from the above literature survey, by increasing the penetration of EVs and PVs in low voltage (LV) and medium voltage (MV) distribution networks, more attention should be paid to coordination between the TSO and DSO.
If the charging pattern of EVs is locally controlled to satisfy the pre-defined charging plan only, serious technical problems can be seen in the hosting capacity (HC) of distribution systems, such as reverse power flows, voltage violations and line congestion \cite{ali2019voltage}. Also, large amounts of EV charging demand can pull up the peak demand of the network and consequently imposes high values of investments for reinforcement of transmission networks as well as power plants.

From the TSO perspective, the heavy loaded distribution feeders push the system toward its loadability limit. Hence, the boundary nodes at the interface of TSO and DSO may be exposed to local voltage instability as a result of vanished available transfer capacity (ATC) of the transmission network as well as uncoordinated operation of local voltage controllers such as OLTCs. In the following section, the research gap that this paper intends to fill is described in more details.
\subsection{Contributions}
In this paper the flexibility resources are optimally dispatched to maximize the EVs and PVs level that can be connected to the distribution network considering the following features:
\begin{itemize}
    \item \textcolor{black}{The uncertainties of PV and EV are taken into account using a scenario-based approach.}
    \item  \textcolor{black}{TSO-DSO interactions is characterized via the driving point impedance concept.
    \item The ZIP models of demands at MV distribution level is considered.
    \item Coordinated operation of the boundary OLTC transformer (HV network's impact on the MV network), PVs, and EVs to provide the volt-var requirements of distribution network, as well as preserve a required level of ATC from TSO's perspective.}
\end{itemize}

\subsection{Paper organization}
The remainder of this paper is organized as follows: 
Section \ref{tsodso} described the TSO-DSO interactions at the presence of DERs. Section \ref{descrip} describes the proposed model for scheduling of EVs via EVCs demand profile modification at the presence of PVs and TSO's considerations. The simulation results are carried out in Section \ref{sec:simulation}. Finally, the conclusions are outlined in Section \ref{sec:Conclusion}.

\section{TSO-DSO interactions at the presence of DERs}
\label{tsodso}
\textcolor{black}{The interaction between upstream HV and downstream MV networks is not negligible at high levels of DERs penetration.
The power and voltage of downstream distribution feeders are supplied via the upstream HV network, which is not an ideal voltage source for both power and voltage. It means that the demand of downstream MV network influences both the voltage and transferable power from the HV network to the MV network. Besides, OLTCs (as the HV to MV networks coupling equipment), play a crucial role, both from TSO and DSO viewpoints. OLTCs keep the voltage of their MV bus in a pre-defined value, by automatic tap adjustment.}

\begin{figure}[!ht]
    \begin{center}
          \centering
    \includegraphics[width=1.1\columnwidth]{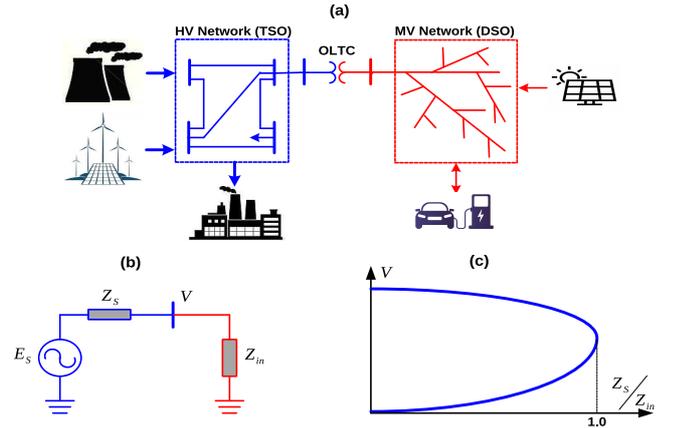}
        \caption{\textcolor{black}{TSO and DSO interaction via OTLCs, (a) conceptual representation, (b) Thévenin network at the HV side, (c) P-V curve.}}
        \label{Fig:conceptual}
    \end{center}
\end{figure}

\textcolor{black}{ The driving point impedance seen from OLTC's HV side depends to the tap ratio and the demand of the downstream network. This impedance determines the available power transfer capacity of the HV network, as depicted in Fig. \ref{Fig:conceptual} which shows the connection of HV and MV networks via an OLTC. On the downstream side, DERs are connected to the distribution feeder. At the HV side of the OLTC, the Thévenin circuit for the rest of the HV network (i.e. $E_{S}$ and $Z_S$) as well as the driving point impedance, i.e. $Z_{in}$, are represented in Fig. \ref{Fig:conceptual}-(b). The ratio of $Z_{S}/Z_{in}$ is the key factor for the determination of ATC (or network loadability) as well as steady-state voltage stability. For a given source impedance, i.e. $Z_S$, by increasing the demand of the downstream network, the voltage at LV side of OLTC reduces. OLTC acts to restore the voltage by increasing its tap ratio. For operation points at the upper part of P-V curve, the impedance seen from the HV side of the OLTC reduces as a result of tap adjustment. By reduction of $Z_{in}$, the ratio of $Z_{S}/Z_{in}$ increases and it tends to $1.00$ at critical operation point, as shown in Fig. \ref{Fig:conceptual}-(c). Thus, the magnitude of $Z_{in}$ is an important factor from the TSO viewpoint, as the tap adjustment by OLTC is crucial for DSO. Hence, coordination between the OLTC operation and the DER optimal scheduling, by considering their impact on the HV network (via $Z_{in}$ property), is a key aspect that this paper aims to study. In the following section, an optimization model is developed for optimal daily scheduling of DERs, by including the concept of driving point impedance as a key constraint.}

\section{DER Scheduling Problem Formulation}
\label{descrip}

In systems with high penetration of EVs, optimization of their charging profile is necessary to ensure that the need for grid development and additional generation capacity is kept at a minimum level. Also, the EV demand should not jeopardize the upstream network's ATC, especially during the heavy loading condition.

By keeping in mind the above considerations, in this section, an optimization model is proposed to maximize the network HC for EVs in the entire scheduling horizon, by modifying the EVCs' charging pattern. Also, without loss of generality of the concept, PVs and EVs are considered as the available DERs in the MV distribution network. The interactions between TSO and DSO in terms of OLTC's tap adjustment as well as the impedance seen through the HV side of the OLTC transformer feeding the MV distribution network, are also considered. 

\textcolor{black}{
\subsection{Uncertainty handling}
Both EVs and PVs are subjected to inherent uncertainties. Additionally, the network demand is also an uncertain parameter \cite{9064904}. These uncertain parameters represent a significant impact on the distribution network scheduling, as they are characterized by a high degree of uncertainty either due to solar irradiance variation, EV owners' behaviors as well as the network demand random variations.\\
\indent For the sake of characterization of the impact of these uncertainties, a scenario-based stochastic optimization technique is developed in this section by considering a set of possible uncertainty scenarios for EVs charging pattern, PV and demand profiles. These scenarios are generated around the mean predicted (or available) profile for EVCs, PVs, and network demand. The model aims to determine an optimal charging profile for both residential and commercial EVCs, considering the concerns of TSO via the aforementioned concept of driving point impedance. The detailed formulation is presented in the following section}

\subsection{Mathematical formulation} 
\label{sec:Formulation}
\subsubsection{Objective function}
\textcolor{black}{At each time step, the capability of distribution network for hosting the EV changes. The aim is to maximize the lower bound of the network HC for EVs, denoted by $\Lambda$, as follows.}
\setlength{\arraycolsep}{0.0em}

\begin{alignat}{2}
\label{TCt}
& \max \ \  \Lambda \\
& Subject \textit { to  } : \notag \\ 
& \eqref{lf1} \textit { to  }\eqref{oltc2}\notag 
\end{alignat}
\textcolor{black}{where $\Lambda$ is the distribution network's minimum HC for EVs.}
\setlength{\arraycolsep}{5pt}
\subsubsection{Problem constraints}
\begin{itemize}
    \item Electricity network constraints:
\end{itemize}
\textcolor{black}{The AC power flow equations are considered as follows ($\forall s \in \Omega_{S}$).}
\begin{alignat}{2}
\label{lf1}
& PG_{b,t,s}+PV_{b,t,s}-PD_{b,t,s}-\sum_{e \in \Omega_{EV}}PE^{e}_{b,t}=\\
&\hspace{7cm}  \sum_{k \in \Omega_{B_b}}{P_{bk,t,s}} \notag \\
\label{lf2}
& QG_{b,t,s}+QV_{b,t,s}-QD_{b,t,s}-\sum_{e \in \Omega_{EV}}QE^{e}_{b,t,s}=\\
&\hspace{7cm}  \sum_{k \in \Omega_{B_b}}{Q_{bk,t,s}} \notag \\
\label{lf1_1}
& PD_{b,t,s}= \\
&\hat{PD}_{b,t,s}\left(KP_{Z,b}(\frac{V_{b,t,s}}{\hat{V}_{b,t,s}} )^2+KP_{I,b}(\frac{V_{b,t,s}}{\hat{V}_{b,t,s}})+KP_{P,b} \right) \notag \\
\label{lf1_2}
& QD_{b,t,s}=\\ 
& \hat{QD}_{b,t,s} \left(KQ_{Z,b}(\frac{V_{b,t,s}}{\hat{V}_{b,t,s}} )^2+KQ_{I,b}(\frac{V_{b,t,s}}{\hat{V}_{b,t,s}})+KQ_{P,b}\right) \notag \\
\label{lf3}
& P_{bk,t,s}=\\
& g_{bk}V^{2}_{b,t,s}-V_{b,t,s}V_{k,t,s} \left( g_{bk}cos(\theta_{bk,t,s}) + b_{bk}sin(\theta_{bk,t,s}) \right) \notag \\
\label{lf4}
& Q_{bk,t,s}=\\ 
& -b_{bk}V^{2}_{b,t,s}- V_{b,t,s}V_{k,t,s} \left(g_{bk}sin\left(\theta_{bk,t,s} \right) - b_{bk}cos\left(\theta_{bk,t,s}\right)\right) \notag \\
\label{lf5}
&\left(P_{bk,t,s}\right)^2+\left( Q_{bk,t,s} \right)^2=\left(V_{b,t,s} I_{bk,t,s}\right)^2\\
\label{lf6}
& \underline{I}_{bk} \le I_{bk,t,s} \le  \overline{I}_{bk} \ \ \ \ \ \forall b \in \Omega_{B}, \forall k \in \Omega_{B_b}\\
\label{lf7}
& \underline{PG}_{b} \le PG_{b,t,s} \le  \overline{PG}_{b}   \ \ \ \ \ \  \forall b \in \Omega_{B_S}\\
\label{lf8}
& \underline{QG}_{b,t} \le QG_{b,t,s} \le  \overline{QG}_{b,t}  \ \ \ \ \ \  \forall b \in \Omega_{B_S}\\
\label{lf9}
& \underline{V}_{b} \le V_{b,t,s} \le  \overline{V}_{b} \ \ \ \ \ \  \forall b \in \Omega_{B}
\end{alignat} 
where \eqref{lf1} and \eqref{lf2} are hourly nodal active and reactive power balances. \textcolor{black}{$PG_{b,t,s}$ and $QG_{b,t,s}$ are active and reactive power generation in node $b$ at time $t$ and scenario $s$, whereas $PV_{b,t,s}$ and $QV_{b,t,s}$ are active power injection and reactive power exchange by the PVs in node $b$ at time $t$ and scenario $s$. Also, $PE^{e}_{b,t}$ and $QE^{e}_{b,t,s}$ are the active power demand of the EV type $e$ connected to bus $b$ at time $t$ and reactive power exchange of the EV type $e$ connected to bus $b$ at time $t$ and scenario $s$, respectively. Besides, $P_{bk,t,s}$ and $Q_{bk,t,s}$ are active and reactive power flowing through the line connecting buses $b$ and $k$ at time $t$ and scenario $s$, respectively.}

Since most of the loads in MV and LV distribution networks are voltage dependent, the ZIP models of active and reactive demands are considered in this paper, as described in \eqref{lf1_1} and \eqref{lf1_2}, respectively. \textcolor{black}{$PD_{b,t,s}$ and $QD_{b,t,s}$ are active and reactive demand in node $b$ at time $t$ and scenario $s$, respectively.} Also, \eqref{lf3}-\eqref{lf5} are the hourly active, reactive, and current flow of the line connecting buses $b$ and $k$ at time $t$ and scenario $s$, whereas its current flow limit is calculated by \eqref{lf6}. \textcolor{black}{$V_{b,t,s}$ and $\theta_{b,t,s}$ are voltage magnitude and angle of node $b$ at time $t$ and scenario $s$, respectively. Besides, $I_{bk,t,s}$ is the current flowing through the line connecting buses $b$ and $k$ at time $t$ and scenario $s$.}

Also, \eqref{lf7} and \eqref{lf8} are the upper/lower limits of active/reactive power exchange with the upstream network, and \eqref{lf9} is the nodal voltage limits, respectively. 

\begin{figure}[!ht]
    \begin{center}
          \centering
\includegraphics[width=0.95\columnwidth, clip=true]{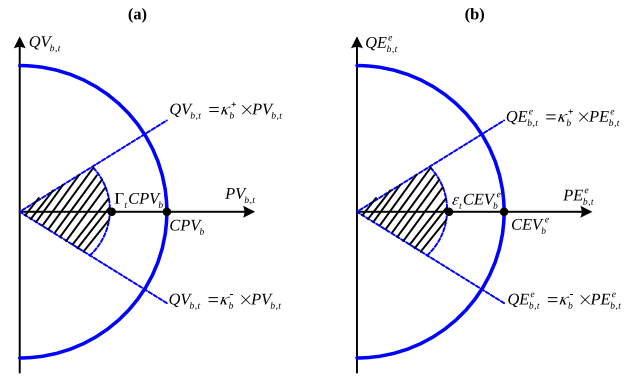}
        \caption{\textcolor{black}{Operation characteristics of PVs and EVs: (a) PV capability curve; (b) EV capability curve.}}
        \label{Fig:Rcapability}
    \end{center}
\end{figure}

\begin{itemize}
    \item PV and EV capability constraints:
\end{itemize}
\textcolor{black}{The DERs such as PVs and EVs should be able to provide reactive power support in order to provide voltage support at the point of power injection \cite{8332112,LI2021123625}. This capability is shown in Fig. \ref{Fig:Rcapability} for PVs and EVs. As it is observed from Fig. \ref{Fig:Rcapability}-(a), for a given amount of active power injection by the PV unit, it can inject or absorb reactive power in a specific level, limited by a minimum power factor. A similar capability curve could be considered for EVs, as shown in Fig. \ref{Fig:Rcapability}-(b). These characteristics can be formulated as follows ($\forall s \in \Omega_{S}$). }

\begin{alignat}{2}
\label{cap1}
& PV_{b,t,s} \le \Gamma_{t,s} \times CPV_{b}   \ \ \ \ \ \  \forall b \in \Omega_{B_{PV}}\\
\label{cap2}
& \left(PV_{b,t,s}\right)^2+\left(QV_{b,t,s}\right)^2 \le  \left(CPV_{b}\right)^2   \ \ \ \ \ \  \forall b \in \Omega_{B_{PV}}\\
\label{cap3}
& \kappa_{PV}^{-} \times PV_{b,t,s} \le QV_{b,t,s} \le \kappa_{PV}^{+} \times PV_{b,t,s}  \ \ \  \forall b \in \Omega_{B_{PV}}\\
\label{cap4}
&0\le PE^{e}_{b,t} \le \varepsilon^{e}_{t} \times CEV^{e}_{b} \\
\label{cap5}
& \left(PE^{e}_{b,t}\right)^2+\left(QE^{e}_{b,t,s}\right)^2 \le  \left(CEV^{e}_{b}\right)^2   \ \ \ \ \ \  \forall b \in \Omega_{B_{EV}}\\
\label{cap6}
& \kappa_{EV}^{-} \times PE^{e}_{b,t} \le QE^{e}_{b,t,s} \le \kappa_{EV}^{+} \times PE^{e}_{b,t}    \ \ \ \  \forall b \in \Omega_{B_{EV}}
\end{alignat}
\textcolor{black}{The hourly PV injection is given by \eqref{cap1}. Also, based on Fig. \ref{Fig:Rcapability}, the PVs active and reactive power capability curve is implemented in \eqref{cap2} and \eqref{cap3}. Similarly, the hourly demand of EVCs is expressed in \eqref{cap4}. Also, the capability curve of EVCs is given by \eqref{cap5} and \eqref{cap6}. }

\begin{itemize}
    \item EVs' demand profile flexibility:
\end{itemize}
The aim is to modify the existing demand profile of EVCs, such that as much as possible, more EV energy can be supplied through the existing network. Hence, the existing EV profile, i.e. $\hat{\Xi}^{e}_{t,s}$, is modified to a new profile, i.e. $\varepsilon^{e}_{t}$, as follows ($\forall t \in \Omega_{T}$, $ \forall e \in \Omega_{EV}$, $\forall b \in \Omega_{B_{EV}}$ and $\forall s \in \Omega_{S}$).
\begin{alignat}{3}
\label{ev1}
& \sum_{t}{\varepsilon^{e}_{t}} = \sum_{t}\hat{\Xi}^{e}_{t,s}  \\
\label{ev2}
&0\le \varepsilon^{e}_{t} \le 1  
\end{alignat}
Equation \eqref{ev1} ensures that the energy demand of EVCs will remain constant in the entire horizon and \eqref{ev2} limits the modified profile of EVCs. \textcolor{black}{Also, as it can be observed from \eqref{ev1} and \eqref{ev2}, the $\varepsilon^{e}_{t}$ takes the same value in all uncertainty scenarios, as it is not defined on $s$. This means that the model gives a unique charging profile for EVCs in all uncertainty scenarios.\\ \\}

\begin{itemize}
    \item TSO-DSO interactions:
\end{itemize}

\textcolor{black}{The aim is to maximize the lower limit of hourly HC by \eqref{TCt}, such that the impedance seen from the HV side of the OLTC connecting HV and MV networks (i.e. $Z_{in,t,s}$) remains higher than a predefined impedance level (i.e. $Z_{min}$) by a scale factor $\alpha$, as follows.
\begin{alignat}{2}
\label{hc1}
&\Lambda \le \sum_{b \in \Omega_{B_{EV}}} \sum_{e \in \Omega_{EV}}PE^{e}_{b,t} \\
\label{hc2}
&Z_{in,t,s} \ge Z_{min} \times \left(1+ \alpha \right)
\end{alignat}
 By maximizing $\Lambda$ in \eqref{TCt}, one can obtain the highest level of the network's HC that is ensured in the entire scheduling horizon for all uncertainty scenarios. \textcolor{black}{Also, as shown in Fig. \ref{Fig:69bus}, $Z_{in,t,s}$ in \eqref{hc2} is the impedance seen through the HV side of the OLTC at time step $t$ and scenario $s$}. This impedance reflects the impact of the downstream MV network on the HV network ATC as well as steady state voltage stability. Besides, the OLTC automatic voltage control capability ensures a regulated voltage at its MV terminal, as follows ($\forall s \in \Omega_{\Omega_{S}}$).
\begin{alignat}{2}
\label{oltc1}
& V_{b,t,s}=V^{spc}_{b,t} \ \ \ \ \ \ \ \ \forall b \in \Omega_{B_{S}} \\
\label{oltc2}
& \underline{T} \le T_{t,s} \le \overline{T}
\end{alignat}
}

\subsection{ The proposed procedure for EV demand profile modification}
\textcolor{black}{For the sake of determining modified profiles of EVCs based on the aforementioned formulation, it is necessary to determine $Z_{min}$ in \eqref{hc2} at first. Actually, $Z_{min}$ is the minimum value of impedance seen from the HV side of the OLTC. For its determination, one can easily solve the above model by excluding Eq. \eqref{hc2} and the minimum value of obtained $Z_{in,t,s}$ is $Z_{min}$. Then, the optimal profile of EVCs could be determined for each level of impedance scale factor $\alpha$.}

\section{Numerical Studies}
\label{sec:simulation}
\subsection{Data}
\begin{figure}[!ht]
    \begin{center}
          \centering
\includegraphics[width=0.95\columnwidth, clip=true]{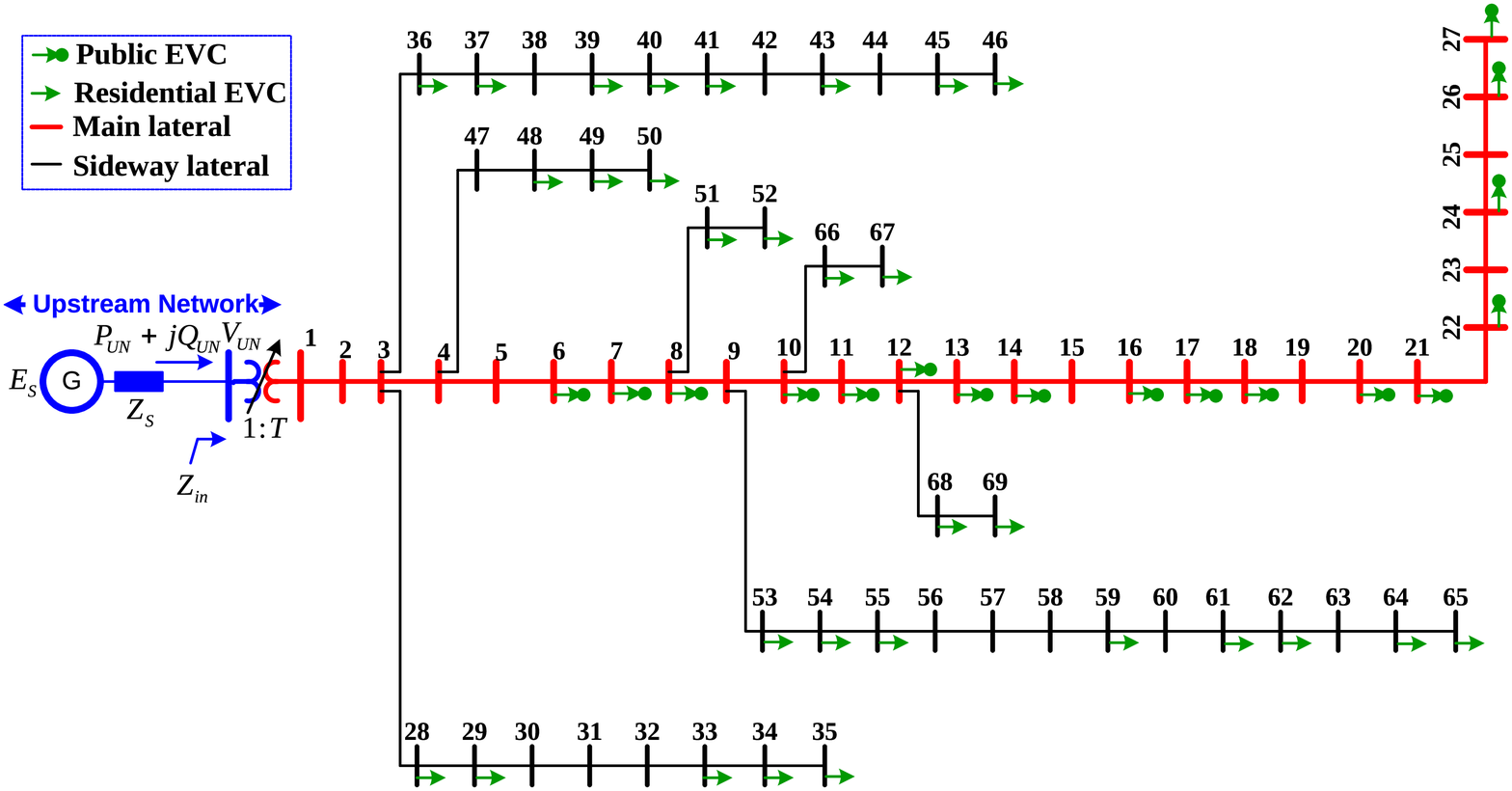}
        \caption{Single-line diagram of the IEEE 69-bus feeder.}
        \label{Fig:69bus}
    \end{center}
\end{figure}

\begin{figure}[!ht]
    \begin{center}
          \centering
    \includegraphics[width=1\columnwidth]{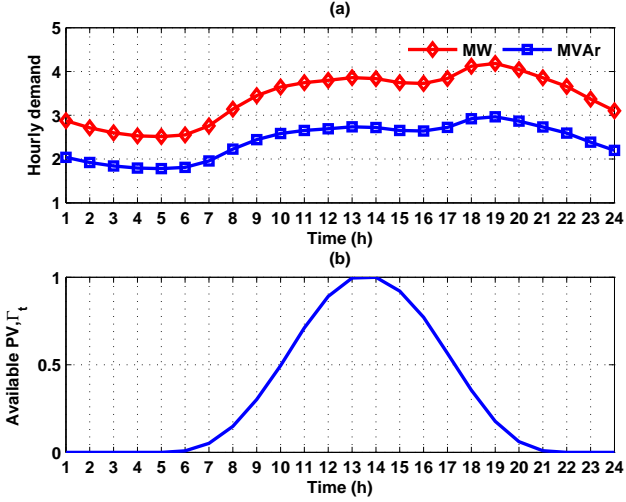}
        \caption{\textcolor{black}{Hourly demand and PV data, (a) Nominal active/reactive power demand, (b) Hourly profile of PVs.}}
        \label{Fig:Hdem}
    \end{center}
\end{figure}

In order to obtain optimal demand pattern for EVs, numerical studies are done on the IEEE 69-bus standard MV distribution network using GAMS optimization package \cite{Soroudi2017}. The single-line diagram of this system is depicted in Fig. \ref{Fig:69bus}. Also, the network data is adopted from \cite{zimmerman2010matpower}. The thermal current limit of the feeder is assumed to be $350A$ which is the same for all line segments. The rated capacity of the OLTC transformer connecting HV and MV networks is assumed to be $10 MVA$. This capacity implies that the limits on the active and reactive power absorbed from the upstream network are $10MW$ and $10MVAr$, respectively. Also, the upper/lower limits of voltage and OLTC's tap are assumed to be $1.100/0.900 \ pu$. \textcolor{black}{A daily 24 hours scheduling horizon is considered with a 1-hour resolution} and the nominal hourly active and reactive power demand of the network is shown in Fig. \ref{Fig:Hdem}, which is distributed in the network based on the percentages given in \cite{zimmerman2010matpower}. Since the network demand is assumed to be voltage-dependent, this nominal demand corresponds to nominal voltage, i.e. $\hat{V}_{b,t,s}=1.00 \ pu$. It is observed from Fig. \ref{Fig:Hdem} that the critical demand interval is $t_{10}-t_{22}$ with the peak demand at hour $t_{19}$.
Also, it is assumed that $65\%$, $20\%$ and $15\%$ of the demand in each load point are constant impedance, constant current, and constant power load, respectively. Besides, as shown in Fig. \ref{Fig:conceptual}-(b), it is assumed that for the entire scheduling horizon, $E_S=1.000 \ pu$ and $Z_S=0.001+j0.007 \ pu$, though one can easily consider them as time-dependent parameters.
 
The aim is to determine the optimal EVs' profile in residential and public (or commercial) sectors. The residential EVCs type is assumed to be \textit{Standard Type-2 (ST2)}, whereas the fast chargers (\textit{AC43, CCS, and CHAdeMO}) are commonly employed in public locations because of their fast charging capabilities. Both charger technologies are commonly used in EU countries. It is assumed that the public EVCs are located on the main lateral (backbone feeder as shown in Fig. \ref{Fig:69bus}), each charge point with the aggregated capacity of $6\times50kW$. Furthermore, the residential EVCs are located in the spurs, each point with an aggregated capacity of $15\times22kW$. It is worth to mention that the typical rating of fast public chargers is $50 kW$ (as already for \textit{CCS} and \textit{CHAdeMO} technologies), whereas, for the common residential three-phase chargers (e.g. \textit{ST2} technology) it is $22 kW$. 

\begin{figure}[!ht]
    \begin{center}
          \centering
    \includegraphics[width=1\columnwidth,]{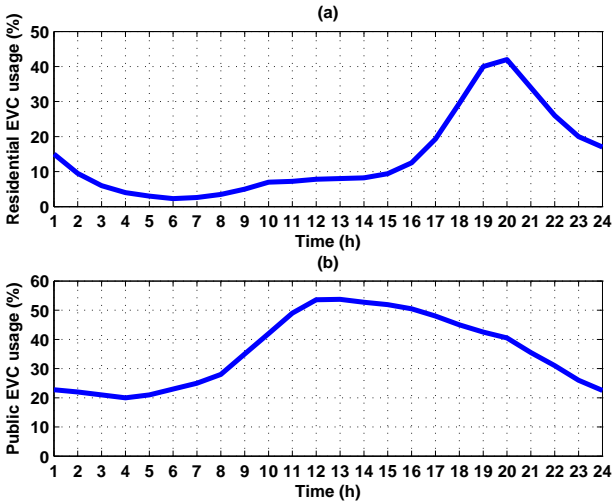}
        \caption{\textcolor{black}{Existing EVC patterns, (a): Residential EVCs \cite{corliss2020electric}, (b): Public EVCs \cite{pallonetto2020framework}}}
        \label{Fig:EVPROF}
    \end{center}
\end{figure}

The existing average demand profiles (i.e. average of $\hat{\Xi}^{e}_{t,s}$) of residential and public EVCs are depicted in Fig. \ref{Fig:EVPROF} which are adopted from \cite{corliss2020electric} and \cite{pallonetto2020framework}, respectively. These profiles which have been obtained through a large data set of EVs charge events during the years 2017 to 2019 in UK and Ireland show the average daily usage of each EVC type. It can be observed from Fig. \ref{Fig:EVPROF}-(a) that in the baseline EVCs' profiles, the residential EVs' demand is mainly imposed during the early night hours, coinciding with the system demand peak interval. In contrast, as it is evident from Fig. \ref{Fig:EVPROF}-(b), the public EV chargers are mainly demanded during the daytime. Based on the system demand profile shown in Fig. \ref{Fig:Hdem}-(a), these types of EVC demand behavior impose more stress on the existing electricity network as the EV demand is added during the system peak interval. Hence, in order to minimize the need for grid development (i.e. more investments on new assets) and the addition of generation capacity, it is necessary to modify the EVs' charging demand pattern.

The PV data is adopted from UK database for the interval of 1/1/2020 to 26/6/2020 \cite{G5}. The mean PV profile (i.e. $\Gamma_{t,s}$ in \eqref{cap1}) that is normalized based on its peak injection is depicted in Fig. \ref{Fig:Hdem}-(b). Also, it is assumed that $20\times2.3 kW$ aggregated PV generation capacities are installed at each load point as shown in Fig. \ref{Fig:69bus}. 

\textcolor{black}{
In order to characterize the impacts of uncertain parameters such as PV power generation, EVCs' charging behaviour, and the demand of network, two cases are studies as follows:}
\begin{itemize}
    \item \textcolor{black}{\textit{Case-I}: deterministic model. In this case, the minimum HC of the distribution network is determined for the mean value of above uncertain parameters. The mean values for active/reactive power demand, PVs' available power generation and EVCs' charging profile are given in Figs. \ref{Fig:Hdem} and \ref{Fig:EVPROF}, respectively. Also, in this case, various sensitivity analyses are carried out to clarify the features of the proposed concept.
    \item \textit{Case-II}: stochastic model. In this case, the uncertainties are handled via a scenario-based stochastic programming technique. Five different scenarios are considered around the aforementioned mean values of the uncertain parameters outlined in \textit{Case-I}.}
\end{itemize}

\textcolor{black}{
\subsection{Case-I: deterministic model}}
\textcolor{black}{The first step is to determine $Z_{min}$ which is the impedance seen from the HV side of OLTC when the constraint \eqref{hc2} is neglected in the scheduling of EVs and PVs. For the studied network, this impedance is obtained as $1.269 pu$ which corresponds to the demand peak hour, i.e. $t_{19}$. Also, the lower bound of HC (i.e. $\Lambda$) in this case is $2.674 MW$.}

\begin{figure}[!ht]
    \begin{center}
          \centering
    \includegraphics[width=1.07\columnwidth,]{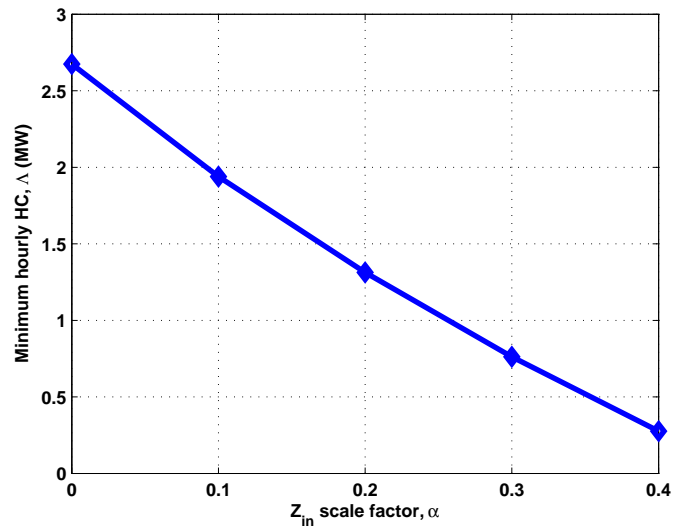}
        \caption{\textcolor{black}{Variation of minimum hourly HC vs $\alpha$}}
        \label{Fig:HCvsAlpha}
    \end{center}
\end{figure}

\textcolor{black}{Based on the obtained value for $Z_{min}$, the model is solved for various levels of impedance scale factor, i.e. $\alpha$. Figure \ref{Fig:HCvsAlpha} shows the variation of $\Lambda$ versus different $\alpha$ levels. It is inferred from this figure that considering TSO's concern (i.e., Eq. \eqref{hc2}) in the model, limits the HC of downstream MV network considerably, such that the HC decreases from $2.674MW$ for $\alpha=0$ to $0.276MW$ for $\alpha=0.40$.}

\begin{figure}[!ht]
    \begin{center}
          \centering
    \includegraphics[width=1.05\columnwidth]{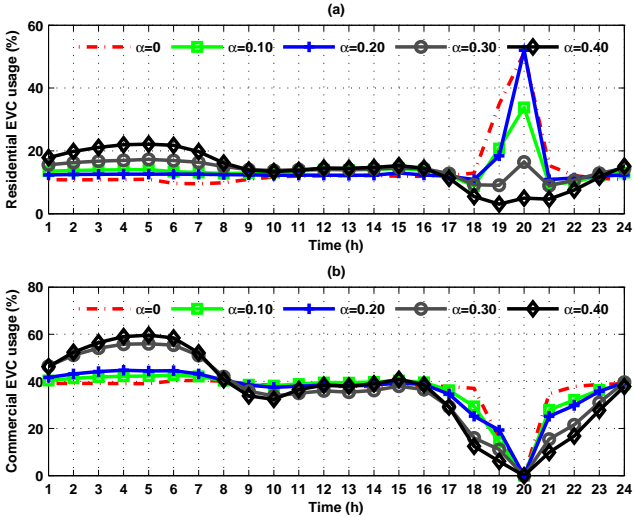}
        \caption{\textcolor{black}{Hourly EVCs' profiles for different values of $\alpha$; (a): Residential EVCs, (b): Public EVCs.}}
        \label{Fig:EVprofs}
    \end{center}
\end{figure}

\textcolor{black}{Also, the hourly EVCs profiles for both residential and commercial EVCs are given in Fig. \ref{Fig:EVprofs}. 
In Fig. \ref{Fig:EVprofs}-(a) the modified hourly profiles of residential EVCs are depicted. By comparing this figure and Fig. \ref{Fig:EVPROF}-(a), it is observed that imposing impedance constraint, i.e. \eqref{hc2}, reduces the EV's demand in the interval $t_{18}-t_{21}$ and the demand is shifted mainly to the light loading condition, i.e. the interval $t_1-t_8$. Similar behavior exists for commercial EVC, as shown in Fig. \ref{Fig:EVPROF}-(b). It is observed that at the presence of PVs power injection, more EVCs demand could be supplied during the daytime, especially in the interval $t_{10}-t_{17}$.}

\begin{figure}[!ht]
    \begin{center}
          \centering
    \includegraphics[width=1.05\columnwidth]{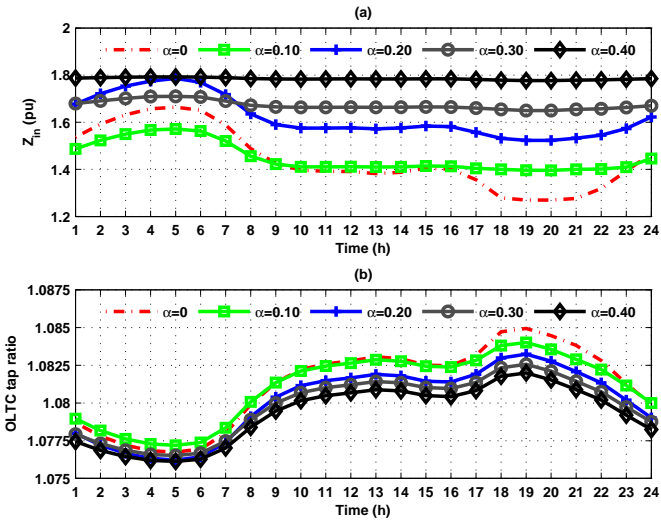}
        \caption{\textcolor{black}{TSO and DSO interactions: (a) Impedance seen from the HV side of OLTC, (b) Tap settings of the OLTC.}}
        \label{Fig:TapZin}
    \end{center}
\end{figure}

\textcolor{black}{The TSO and DSO have interactions in terms of OLTC's tap settings and $Z_{in}$ seen from HV side of the OLTC. This mutual impact of TSO and DSO is shown in Fig. \ref{Fig:TapZin}. It is observed from Fig. \ref{Fig:TapZin}-(a) that for $\alpha=0$, the hourly impedance is limited to $Z_{min}$ via constraint \eqref{hc2} in $t_{19}$ and during light loading condition, impedance is sufficiently higher than $Z_{min}$. By increasing the $\alpha$, variation of $Z_{in,t}$ decreases, such that for $\alpha=0.4$, which is the most conservative level of the impedance seen by the upstream network, an even $Z_{in,t}$ is seen at the entire scheduling horizon. The impedance seen by the upstream network depends on the tap ratio of OLTC, when the OLTC tries to keep its MV bus voltage at a preset value of $V^{spc}_{b,t}=1.060 \ \ pu$ in \eqref{oltc1}. The corresponding hourly tap behavior is depicted in \ref{Fig:TapZin}-(b).}

\begin{figure}[!ht]
    \begin{center}
          \centering
    \includegraphics[width=1.1\columnwidth]{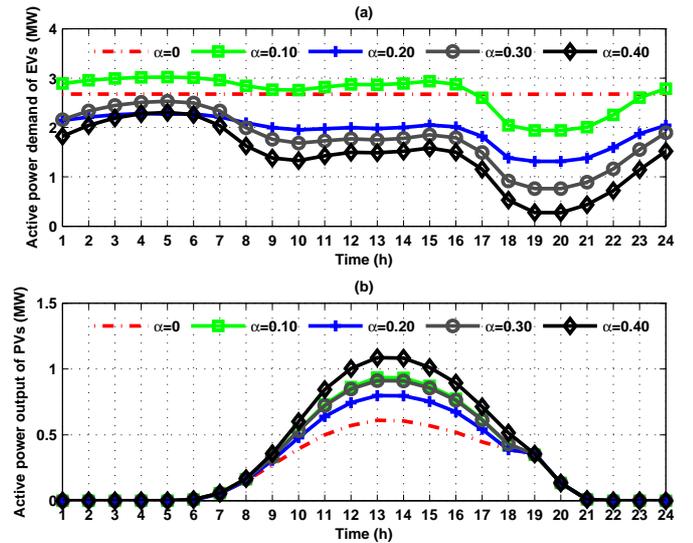}
        \caption{\textcolor{black}{EVs and PVs hourly power schedule for different values of $\alpha$.}}
        \label{Fig:EVPV}
    \end{center}
\end{figure}

\textcolor{black}{As it is observed, introducing the need for consideration of minimum impedance as well as OLTC's tap settings in a coordinated manner, manipulates the EVs and PVs schedules, considerably. The hourly schedules of EVCs and PVs are presented in Fig. \ref{Fig:EVPV}. It is inferred from Fig. \ref{Fig:EVPV}-(a) that at the presence of \eqref{hc2}, the demand incurred by EVCs is limited during the peak demand interval in higher levels of $\alpha$. This is in line with the fact that downstream network demand can have a considerable impact on the ATC of upstream HV network.
Also, it is observed from Fig. \ref{Fig:EVPV}-(b) that power injection by PVs increases versus the $\alpha$ in order to supply more demands during the daytime interval.} 
\textcolor{black}{Besides, it is observed that EVs energy consumption can increase the energy injection via PVs, since the operational limits such as voltage rise and feeder thermal limits can be alleviated thoroughly.}

\begin{figure}[!ht]
    \begin{center}
          \centering
    \includegraphics[width=1.1\columnwidth]{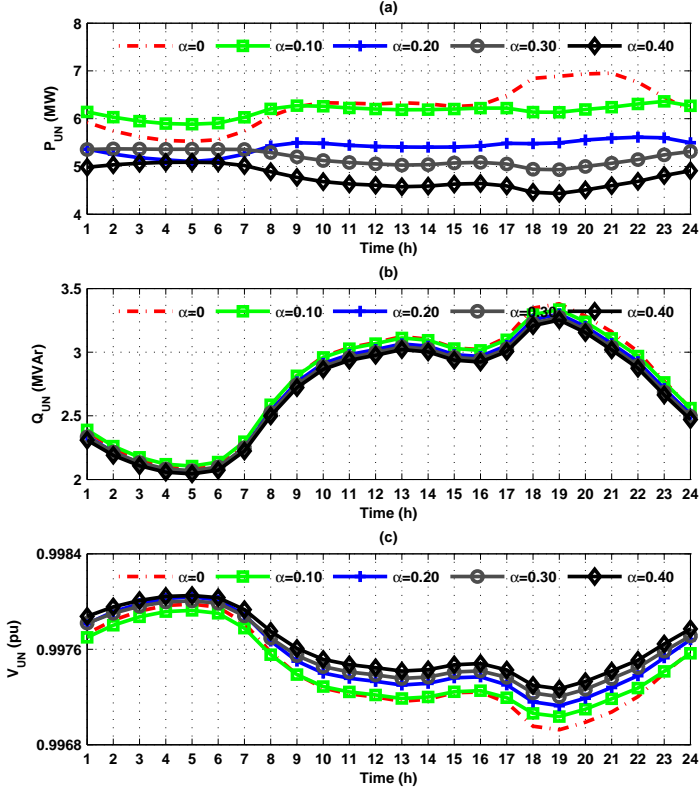}
        \caption{\textcolor{black}{Hourly values of OLTC's HV side variables: (a) Active power incurred, (b) Reactive power incurred, (c) Voltage profile.}}
        \label{Fig:slack_vars}
    \end{center}
\end{figure}

\textcolor{black}{The overall behavior of downstream network seen from the OLTC's HV bus, in terms of active/reactive demand and voltage of HV bus (as also shown in Fig. \ref{Fig:69bus}) is shown in Fig. \ref{Fig:slack_vars}. It is observed from Fig. \ref{Fig:slack_vars}-(a) that the peak demand of downstream network decreases as $\alpha$ increases. Since the EVs and PVs are scheduled in unity power factor, the reactive power demand incurred via HV upstream network is almost the same for all levels of $\alpha$ as shown in  Fig. \ref{Fig:slack_vars}-(b). The hourly voltage profile of OLTC's HV bus for different values of $\alpha$ is also presented in Fig. \ref{Fig:slack_vars}-(c). It is observed from this figure that the voltage highly depends on the reactive power demand of downstream network since a high $X/R$ ratio is assumed for the upstream network's Thévenin impedance, in this study.}

\textcolor{black}{
In the following, some sensitivity analyses and further studies are performed to clarify various features of the proposed model for distribution networks minimum hourly HC maximization.}\\

\subsubsection{\textcolor{black}{The impacts of DERs power factor and OLTC's voltage set-point}}
\begin{figure}[!ht]
    \begin{center}
          \centering
    \includegraphics[width=1\columnwidth]{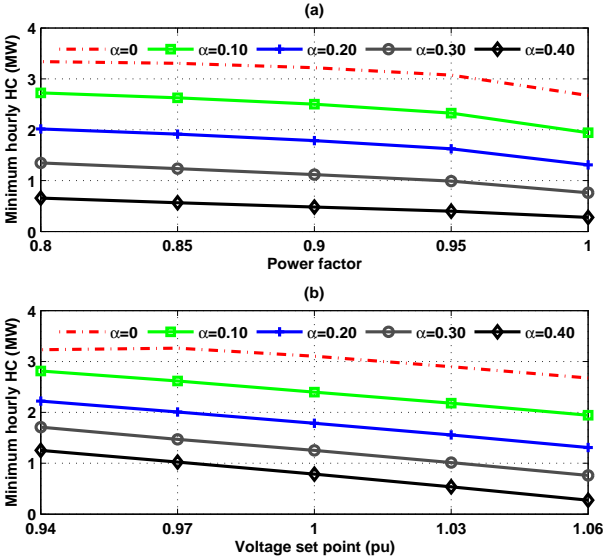}
        \caption{\textcolor{black}{Sensitivity of HC, ($\Lambda$) in different levels of $\alpha$, (a) Vs DERs power factor, (b) Vs, OLTC's voltage set point }}
        \label{Fig:HCminPF}
    \end{center}
\end{figure}

\textcolor{black}{In order to show the impact of DERs power factor as well as the set point voltage of the OLTC on the MV network HC for EVCs, two sensitivity analyzes are performed. As shown in Fig. \ref{Fig:HCminPF}-(a), by increasing the power factor of PVs and EVs from $0.80$ to $1.00$ (that is included in $\kappa_{EV/PV}^{-/+}$ in \eqref{cap3} and \eqref{cap6}), the network's HC decreases, accordingly. This figure implies the role of DERs' reactive power support to increase the HC for different values of $\alpha$.
The impact of voltage setting of OLTC, i.e. $V^{spc}_{b,t}$ in \eqref{oltc1}, on the HC is also shown in Fig. \ref{Fig:HCminPF}-(b). It is observed from this figure that by increasing the voltage set-point of OLTC, the HC decreases for all levels of $\alpha$, since the higher the voltage set-point the higher demand of MV network as well as the lower impedance seen from the HV side of the OLTC.  }

\subsubsection{\textcolor{black}{The impact of various capacities of EVCs and PVs}}
\begin{figure}[!ht]
    \begin{center}
          \centering
    \includegraphics[width=1.1\columnwidth]{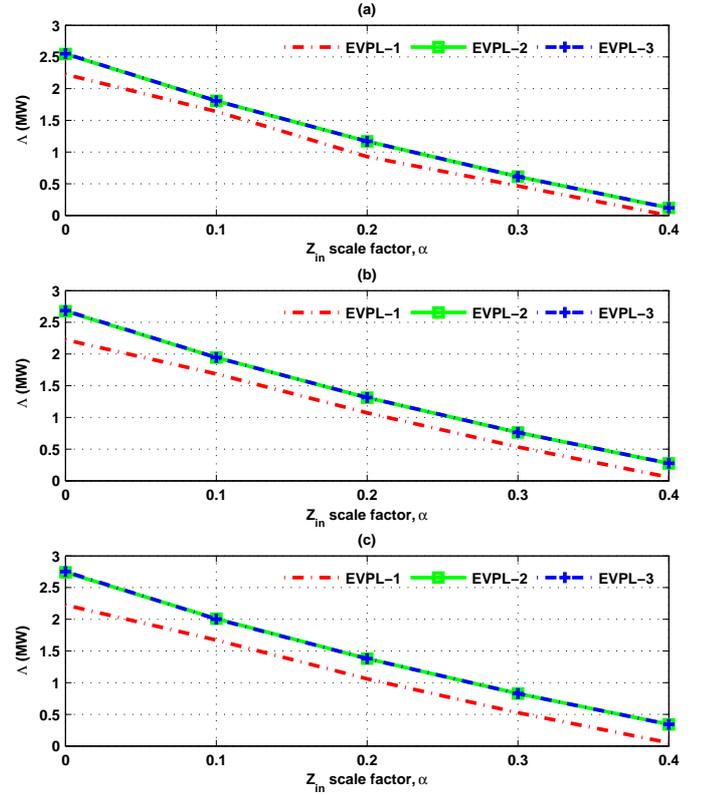}
        \caption{\textcolor{black}{Sensitivity of minimum HC, at different levels of $\alpha$ and EVPLs for: (a) PVPL-1, (b) PVPL-2, (c) PVPL-3 }}
        \label{Fig:HC_Cap_sens}
    \end{center}
\end{figure}

\textcolor{black}{The impact of various capacities of EVs and PVs on the network's minimum HC is studied. In this analysis, three different EV penetration levels (EVPLs) are considered, as follows:
\begin{itemize}
    \item EVPL-1: $4 \times 50kW$ for public EVCs and $10 \times 22kW$ residential EVs.
    \item EVPL-2: $6 \times 50kW$ for public EVCs and $15 \times 22kW$ residential EVs.
    \item EVPL-3: $8 \times 50kW$ for public EVCs and $20 \times 22kW$ residential EVs.
\end{itemize}
Besides, for PVs three different PLs (PVPLs) are considered as follows:
\begin{itemize}
    \item PVPL-1: $10 \times 2.3kW$ PV capacities.
    \item PVPL-2: $20 \times 2.3kW$ PV capacities.
    \item PVPL-3: $30 \times 2.3kW$ PV capacities.
\end{itemize}
By combining these EVPLs and PVPLs, nine different cases are generated and the minimum HC of the network is obtained via the proposed formulation. The results are shown in Fig. \ref{Fig:HC_Cap_sens} for these different cases. The following results could be inferred from this figure:
\begin{itemize}
    \item By increase of PVPL, no significant increase is observed in the network's minimum HC for EVs. This is because of the fact that the minimum HC is limited due to the demand at $t_{19}$. In this hour, no significant PV generation is available as shown in Fig. \ref{Fig:Hdem}-(b).
    \item By increase of EVPL for a given PVPL, the network HC for EVs increases for all levels of $\alpha$. By increase of EVPL from EVPL-1 to EVPL-2, more increase in HC could be attained, while from EVPL-2 to EVPL-3, no significant enhancement is observed for the HC. For higher levels of EVPL, the limit imposed by \eqref{hc2} limits the minimum HC of the network.
\end{itemize}
}

\subsubsection{\textcolor{black}{The impact of upstream network's Thévenin impedance}}
\begin{figure}[!ht]
    \begin{center}
          \centering
    \includegraphics[width=1\columnwidth]{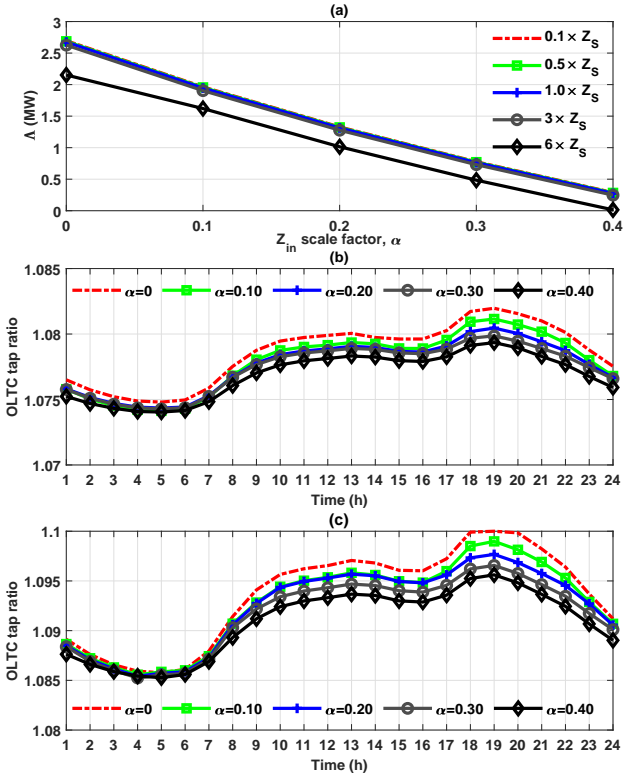}
        \caption{\textcolor{black}{Impact of $Z_S$ on minimum HC: (a) HC versus $\alpha$ for various scales of $Z_S$, (b) OLTC's hourly tap settings for $0.1 \times Z_S$, (c) OLTC's hourly tap settings for $6 \times Z_S$,}}
        \label{Fig:ZS_sens}
    \end{center}
\end{figure}

\textcolor{black}{To show the impact of upstream HV network's Thévenin impedance on the distribution network minimum HC for EVs, another study is performed here, by scaling $Z_S$ from 0.1 to 6 times of the aforementioned value of $0.001+j0.007 pu$.\\
The network minimum HC for different values of $\alpha$ is shown in Fig. \ref{Fig:ZS_sens}-(a). It is inferred from this figure that for 0.1, 0.5 1.0 and 3.0 times of $Z_S$, the HC is not changed considerably. While, for 6.0 times of $Z_S$, the HC decreases significantly, as the ratio of $Z_S/Z_{in}$ is highest for this scale of $Z_S$. For smaller scales of $Z_S$, the highest level of HC could be attained, as the voltage drop on $Z_S$ is not significant and OLTC can regulate its downstream point voltage at $V^{spc}=1.06 pu$. For example, the OLTC's hourly tap adjustment for $0.1 \times Z_S$ is depicted in Fig. \ref{Fig:ZS_sens}-(b). But, when $Z_S$ scales up, the voltage drop across it increases, and to keep the downstream bus voltage in a constant value of $V^{spc}=1.06 pu$ defined in \eqref{oltc1}, the OLTC increases tap ratio, until the tap hits the corresponding upper limit of 1.1 at interval $t_{18}-t_{20}$, as shown in Fig. \ref{Fig:ZS_sens}-(c).}

\subsubsection{\textcolor{black}{The impact of upstream network's variability}}
\begin{figure}[!ht]
    \begin{center}
          \centering
    \includegraphics[width=0.9\columnwidth]{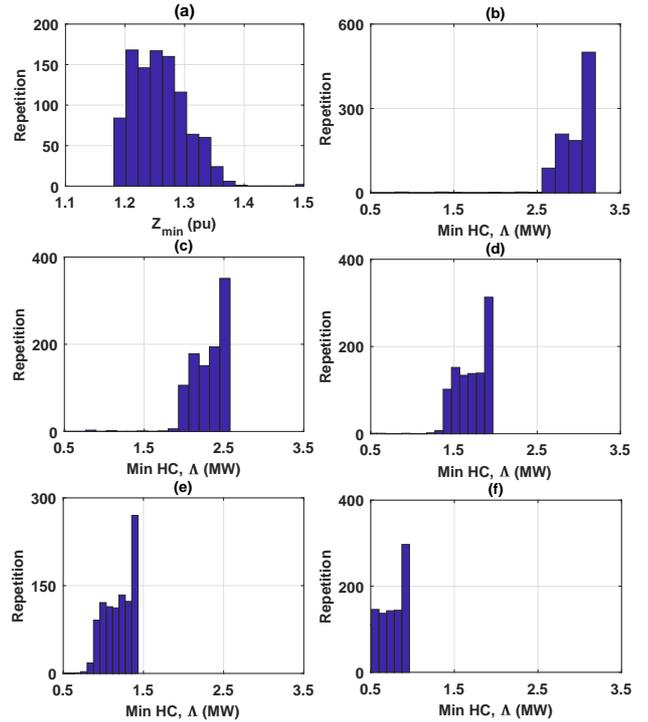}
        \caption{\textcolor{black}{Impact $E_S$ and $Z_S$ randomness on minimum HC: (a) $Z_{min}$ variations, (b) $\Lambda$ variations for $\alpha=0.0$, (c) $\Lambda$ variations for $\alpha=0.1$, (d) $\Lambda$ variations for $\alpha=0.2$, (e) $\Lambda$ variations for $\alpha=0.3$, (f) $\Lambda$ variations for $\alpha=0.4$.}}
        \label{Fig:mcs_zs_es}
    \end{center}
\end{figure}

\textcolor{black}{The impact of variations on the transmission network's topology as well as operation point, can be reflected in $Z_S$ and $E_S$ values. For example, by occurring contingencies such as line, transformer, and generator outage in transmission network, the values of $Z_S$ and $E_S$ change. In such circumstances, $Z_S$ increases as the transmission network tend to weaken subject to the contingencies. Also, $E_S$ varies as a result of variations of the transmission network's operation point, e.g., subject to the contingencies. The impact of $Z_S$ and $E_S$ variations on the proposed model are studied here by performing a Monte-Carlo simulation (MCS)-based analysis. Hence, by applying random variations on the values of $E_S$ and $Z_S$, the network's minimum hourly HC for EVs is calculated by MCS iterations. To characterize the uncertainty of $E_S$ and $Z_S$, uniform randomness is considered for both $E_S$ and $Z_S$. It is assumed that $E_S$ varies in the interval $[0.95 \ \ 1.05]\ \ pu$, uniformly. Also, $Z_S$ varies in the interval $[0.1 \ \ 10.0]$ times of its initial value of $0.001+j0.007 \ pu$. Besides, it is assumed that $V^{spc}_{b,t}=1.000 \ pu$, here. In each iteration of the MCS, random values are obtained for $E_S$ and $Z_S$ in the above intervals. At the beginning of each iteration, $Z_{min}$ value updated based on the $E_S$ and $Z_S$ of that iteration and the MCS run by 1000 iterations.\\
\indent The obtained results are given in Fig. \ref{Fig:mcs_zs_es}. The obtained random values for $Z_{min}$ are shown in Fig. \ref{Fig:mcs_zs_es}-(a). It is inferred from this figure that for the considered interval of uncertainty for $E_S$ and $Z_S$, $Z_{min}$ varies in the range of $[1.2 \ \ 1.4]\ \ pu$. Also, the obtained values for the network's minimum hourly HC for different values of $\alpha$ is depicted in Fig. \ref{Fig:mcs_zs_es}-(b)-(f). It can be seen from these figures that by increase of $\alpha$, the network's HC decreases, such that its mean values for $\alpha=0.0$, $0.1$, $0.2$, $0.3$ and $0.4$, are $2.971 MW$, $2.304 MW$, $1.704 MW$, $1.181 MW$ and $0.723 MW$, respectively. These values can be compared with those values obtained for $E_S=1.00 \ pu$, $Z_S=0.001+j0.007 \ pu$ and $V^{spc}_{b,t}=1.000 \ pu$ in Fig. \ref{Fig:HCminPF}-(b), which are $3.103 MW$, $2.398 MW$,	$1.785 MW$,	$1.254MW$ and $0.786MW$, respectively. One can observe that when the upstream network variability is included, the distribution network's minimum hourly HC averagely decreases by $4.25\%$, $3.91\%$, $4.53\%$, $5.82\%$ and $8.01\%$, respectively. This means that the impact of upstream network's variability on $\Lambda$ increases vs $\alpha$, such that the maximum reduction of $\Lambda$ is obtained for $\alpha=0.40$ which equals to $8.01\%$.}

\begin{figure}[!ht]
    \begin{center}
          \centering
    \includegraphics[width=0.9\columnwidth,]{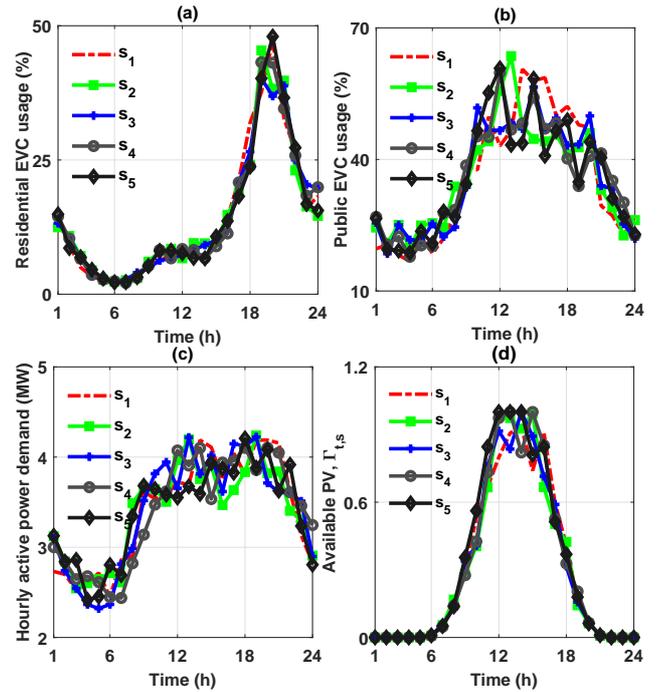}
        \caption{\textcolor{black}{Uncertainty scenarios in \textit{Case-II}: (a) Residential EVCs, (b) Public EVCs, (c) Hourly active power demand, (d) Available PV.}}
        \label{Fig:scens}
    \end{center}
\end{figure}

\begin{figure}[!ht]
    \begin{center}
          \centering
    \includegraphics[width=0.9\columnwidth,]{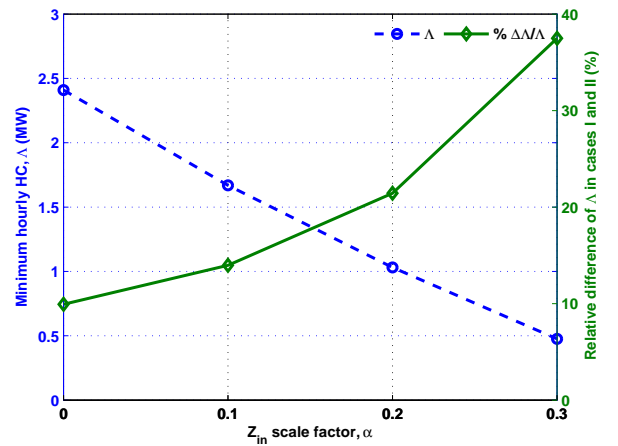}
        \caption{\textcolor{black}{The obtained minimum hourly HC in \textit{Case-II} and its difference with \textit{Case-I}}}
        \label{Fig:hc_scens}
    \end{center}
\end{figure}

\begin{figure}[!ht]
    \begin{center}
          \centering
    \includegraphics[width=1.05\columnwidth,]{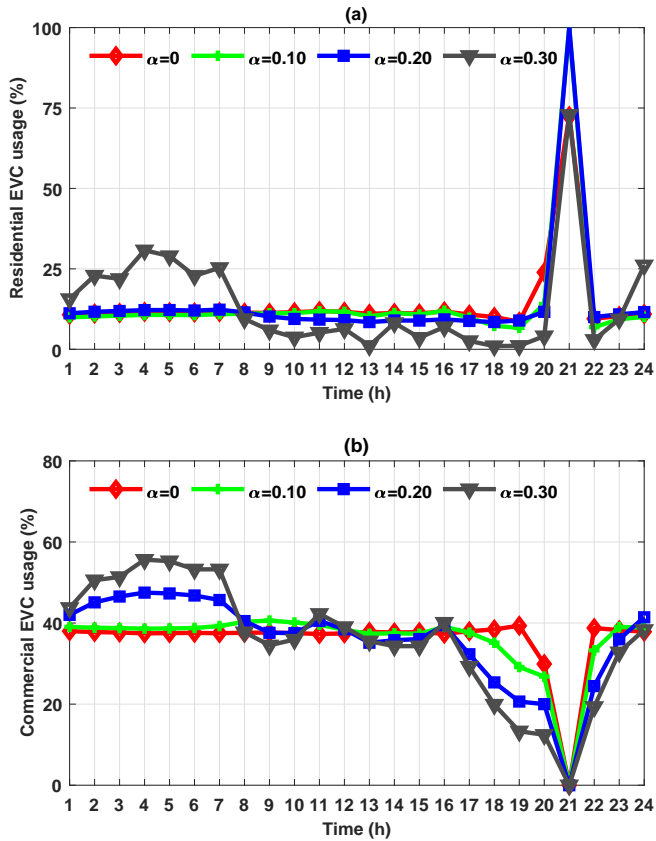}
        \caption{\textcolor{black}{The obtained charging profiles in \textit{Case-II}: (a) Modified profile of residential EVCs, (b) Modified profile of public EVCs.}}
        \label{Fig:charges_scens}
    \end{center}
\end{figure}

\begin{figure}[!ht]
    \begin{center}
          \centering
    \includegraphics[width=0.9\columnwidth,]{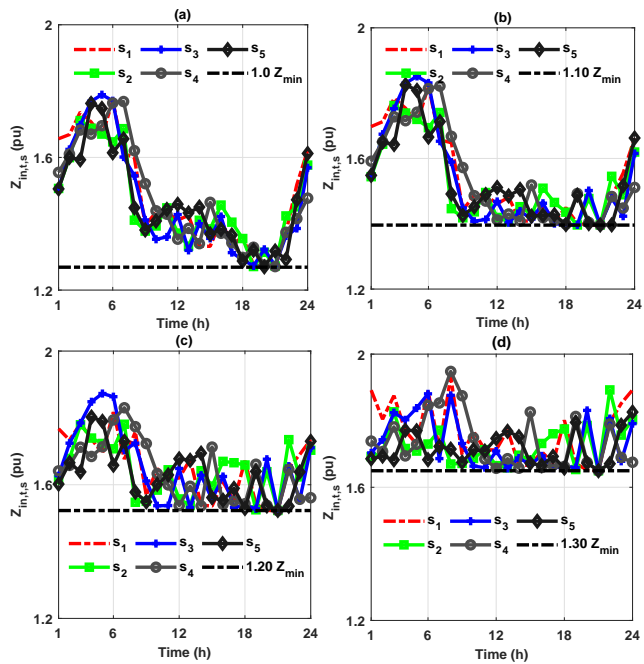}
        \caption{\textcolor{black}{Impedance seen by the HV network, $Z_{in,t,s}$, in different scenarios of \textit{Case-II}: (a) $\alpha=0$, (b) $\alpha=0.1$, (c) $\alpha=0.2$, (d) $\alpha=0.3$.}}
        \label{Fig:zin_scens}
    \end{center}
\end{figure}

\begin{figure}[!ht]
    \begin{center}
          \centering
    \includegraphics[width=0.9\columnwidth,]{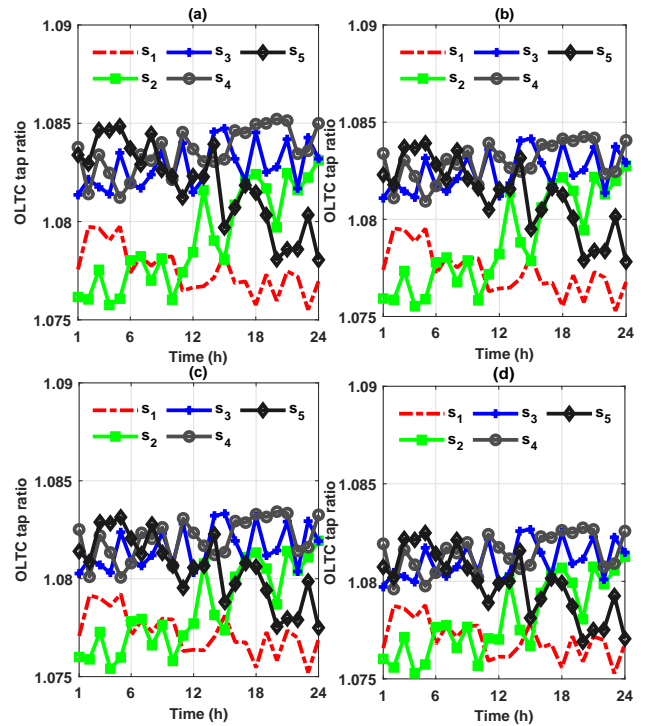}
        \caption{\textcolor{black}{Tap setting of the OLTC in different scenarios of \textit{Case-II}: (a) $\alpha=0$, (b) $\alpha=0.1$, (c) $\alpha=0.2$, (d) $\alpha=0.3$.}}
        \label{Fig:tap_scens}
    \end{center}
\end{figure}

\textcolor{black}{
\subsection{Case-II: stochastic model}}
\textcolor{black}{In this section, uncertainty analysis is performed to address the inherent uncertainties of PVs, EVs, and demand. These uncertain parameters represent a significant impact on the results obtained, as they are characterized by a high degree of uncertainty.\\
\indent As it is aforementioned, to characterize the impact of uncertainties, a scenario-based stochastic optimization technique is deployed by this paper, by considering a set of uncertainty scenarios. Here, five scenarios are generated with the same probabilities based on the mean hourly data of demand, PVs, and EVCs (as given in Figs. \ref{Fig:Hdem} and \ref{Fig:EVPROF}), as follows.
\begin{itemize}
    \item Residential and commercial EVC profiles scenarios are generated by considering a $±20\%$ uniform tolerance around the average profiles given in Fig. \ref{Fig:EVPROF}, respectively. The generated scenarios for residential and public EVCs are given in Fig. \ref{Fig:scens}-(a) and (b), respectively.
    \item Active/reactive power demand scenarios are generated by considering a $±10\%$ uniform tolerance around the average demand profile given in Fig. \ref{Fig:Hdem}-(a). The generated scenarios for active and reactive power demands are assumed to be the same, as the active power demand scenarios are given in Fig. \ref{Fig:scens}-(c). By dividing the demand in each scenario and time step to its mean value, a relative ratio is obtained which is also used for characterizing the reactive power demand scenarios.
    \item PV generation scenarios are generated by considering a $±20\%$ uniform tolerance around the average PV profile given in Fig. \ref{Fig:Hdem}-(b). The generated PV scenarios are given in Fig. \ref{Fig:scens}-(d).
\end{itemize}
By considering these scenarios, the value of $Z_{min}$ which is the minimum impedance seen from the HV side of the OLTC in all scenarios, is $1.268 pu$. Now, by knowing this value of $Z_{min}$, the stochastic model is solved and the network's minimum hourly HC of EVs is obtained for various levels of $\alpha$, as given in Fig. \ref{Fig:hc_scens}. In some scenarios, the EVs and demand of the network are higher than the mean value, such that preserving the desired voltage of $V^{spc}=1.06 pu$ at the MV side of the OLTC is not possible, as the OLTC's tap hits its upper limit. Hence, in this case, impedance scale factor, i.e., $\alpha$, is limited to $0.30$, as shown in Fig. \ref{Fig:hc_scens}.
By comparing the obtained values of $\Lambda$ in this case with those obtained for the mean value of scenarios (depicted in Fig. \ref{Fig:HCvsAlpha}), it can be observed that the network's HC decreases when considering these scenarios. As an example, for $\alpha=0.2$ the minimum hourly HC in \textit{Case-I} is $1.313 MW$, whereas it reduces to $1.031 MW$ in \textit{Case-II} as given in Fig. \ref{Fig:hc_scens}, substantiating a $21.5\%$ reduction in the network's HC for EVs. The amount of $\Lambda$ reduction in \textit{Case-II} with respect to \textit{Case-I} is higher for greater values of $\alpha$.\\
\indent The optimal charging profiles of residential and public EVCs are also depicted in Fig. \ref{Fig:charges_scens}-(a) and (b), respectively. By comparison of these profiles with the corresponding profiles in \textit{Case-I} (given in Fig. \ref{Fig:EVprofs}-(a) and (b)), one can observe that the charging profile obtained for residential EVCs in \textit{Case-II} has a significant value in $t_{21}$ for all values of $\alpha$, which is higher than that of \textit{Case-I}. Also, in \textit{Case-II}, the charging profile is almost the same for $\alpha=0-0.20$, whereas it has a shift from daytime to early morning interval for $\alpha=0.30$.
Besides, almost the same profiles are obtained in cases \textit{I} and \textit{II} for public EVCs, as shown in Fig. \ref{Fig:charges_scens}-(b).\\
\indent The impedance seen at the HV side of the OLTC at different scenarios of \textit{Case-II} is shown in Fig. \ref{Fig:zin_scens} for different values of $\alpha$. It can be observed that by the increase of $\alpha$, the impedance seen by the upstream HV network increases in all scenarios, due to the binding constraint \eqref{hc2}. Also, for all scenarios by the increase of $\alpha$, $Z_{in,s,t}$ hits its minimum limit defined by \eqref{hc2}, more frequently.\\
\indent The hourly tap ratio of the OLTC in different scenarios of \textit{Case-II} for various levels of $\alpha$ is given by Fig. \ref{Fig:tap_scens}. In all cases, the OLTC aims to keep its MV side bus voltage at $V^{spc}=1.06 pu$. It is observed from this figure that for smaller values of $\alpha$, as the EVs demand increases, the tap levels are higher in comparison with those of higher levels of $\alpha$. Besides, for a given level of $\alpha$ in scenarios with higher levels of EVs and demand, a bigger tap ratio is required to keep the voltage at $1.06 pu$.
}

\section{Conclusions}
\label{sec:Conclusion}
In this paper, the TSO and DSO interactions are included in the scheduling of DERs, namely EVs and PVs. Optimal schedules of EVs are determined in a way that the MV network is capable to deliver the energy demand as much as possible while keeping the ATC of the upstream HV network. Also, coordination of OLTC at the boundary bus connecting TSO and DSO is done along with optimal scheduling of EVs and PVs. To preserve a proper ATC in the HV network, the concept of driving point impedance is included in the proposed model. In this way, the DSO can determine optimal schedules of DERs, as well as a preset margin for loadability of upstream HV network ensured. 

The proposed model is implemented on the IEEE 69-bus distribution system, connected to the upstream transmission network via an OLTC. \textcolor{black}{The uncertain nature of PV generation, EV behavior, and the network demand is modeled via a scenario-based stochastic optimization model and detailed results are presented in deterministic and stochastic cases}. The following conclusions can be outlined:

\begin{itemize}
    \item Incorporating the TSO's concerns in the DERs scheduling at distribution network, changes the EVCs demand profile considerably, such that the charging demand reduces during the heavy loading condition.
    \item The voltage set-point of the OLTC is the key parameter that affects the distribution network capability in supplying the EV demand. It is observed that the higher voltage set-point of OLTC is, the lower the distribution network HC will be for EVs.
    \item Reactive power support by the PVs and EVs has a considerable impact on the network's ability to supply more EVs demand. When reactive power is injected into the network by PVs and EVs, the impedance seen from the HV side of the OLTC increases, which creates more ATC.
    \item \textcolor{black}{When the upstream network's variability is considered, the distribution network's minimum hourly HC decreases.}
    \item \textcolor{black}{When the uncertainty of DERs and demand is considered, the network minimum hourly HC for EVs decreases in comparison with the deterministic model. The minimum HC reduction at the presence of uncertainties depends on the impedance scaling factor, such that the greater the impedance scaling factor the higher the reduction of HC in the stochastic case compared to the deterministic case.}
\end{itemize}
\section*{Acknowledgements}
The work done by Abbas Rabiee has been funded [in
part] by the Irish Government Department of Communications, Climate Action, and Environment. The work done by Alireza Soroudi in this publication has emanated from research supported in part by Science Foundation Ireland (SFI) under the SFI Strategic Partnership Programme Grant Number SFI/15/SPP/E3125.

\bibliographystyle{IEEEtran}
\bibliography{MP-ref}

\begin{thebibliography}{10}
\providecommand{\url}[1]{#1}
\csname url@samestyle\endcsname
\providecommand{\newblock}{\relax}
\providecommand{\bibinfo}[2]{#2}
\providecommand{\BIBentrySTDinterwordspacing}{\spaceskip=0pt\relax}
\providecommand{\BIBentryALTinterwordstretchfactor}{4}
\providecommand{\BIBentryALTinterwordspacing}{\spaceskip=\fontdimen2\font plus
\BIBentryALTinterwordstretchfactor\fontdimen3\font minus
  \fontdimen4\font\relax}
\providecommand{\BIBforeignlanguage}[2]{{%
\expandafter\ifx\csname l@#1\endcsname\relax
\typeout{** WARNING: IEEEtran.bst: No hyphenation pattern has been}%
\typeout{** loaded for the language `#1'. Using the pattern for}%
\typeout{** the default language instead.}%
\else
\language=\csname l@#1\endcsname
\fi
#2}}
\providecommand{\BIBdecl}{\relax}
\BIBdecl

\bibitem{iea2020}
IEA, \emph{Global EV Outlook 2020}.\hskip 1em plus 0.5em minus 0.4em\relax IEA,
  Paris, 2020.

\bibitem{mintzer2003us}
I.~Mintzer, J.~A. Leonard, and P.~Schwartz, \emph{US energy scenarios for the
  21st century}.\hskip 1em plus 0.5em minus 0.4em\relax Pew Center on Global
  Climate Change, 2003.

\bibitem{ukfes}
NationalGrid, \emph{Future Energy Scenarios}.\hskip 1em plus 0.5em minus
  0.4em\relax UK National Grid Electricity System Operator (ESO), 2019.

\bibitem{EirGrid}
EirGrid, \emph{Tomorrow’s Energy Scenarios 2019 Ireland}.\hskip 1em plus
  0.5em minus 0.4em\relax EirGrid, Ireland, October, 2019.

\bibitem{6919255}
J.~C. {Mukherjee} and A.~{Gupta}, ``A review of charge scheduling of electric
  vehicles in smart grid,'' \emph{IEEE Systems Journal}, vol.~9, no.~4, pp.
  1541--1553, 2015.

\bibitem{DANESHVAR2020119267}
M.~Daneshvar, B.~Mohammadi-Ivatloo, and K.~Zare, ``Two-stage optimal robust
  scheduling of hybrid energy system considering the demand response
  programs,'' \emph{Journal of Cleaner Production}, vol. 248, p. 119267, 2020.

\bibitem{9064904}
A.~{Ali}, K.~{Mahmoud}, D.~{Raisz}, and M.~{Lehtonen}, ``Probabilistic approach
  for hosting high pv penetration in distribution systems via optimal oversized
  inverter with watt-var functions,'' \emph{IEEE Systems Journal}, vol.~15,
  no.~1, pp. 684--693, 2021.

\bibitem{7491385}
I.~J. {Perez-Arriaga}, ``The transmission of the future: The impact of
  distributed energy resources on the network,'' \emph{IEEE Power and Energy
  Magazine}, vol.~14, no.~4, pp. 41--53, 2016.

\bibitem{gonzalez2020smart}
D.~M. Gonzalez, J.~Myrzik, and C.~Rehtanz, ``The smart power cell concept:
  mastering tso--dso interactions for the secure and efficient operation of
  future power systems,'' \emph{IET Generation, Transmission \& Distribution},
  vol.~14, no.~13, pp. 2407--2418, 2020.

\bibitem{kalantar2019characterizing}
M.~Kalantar-Neyestanaki, F.~Sossan, M.~Bozorg, and R.~Cherkaoui,
  ``Characterizing the reserve provision capability area of active distribution
  networks: A linear robust optimization method,'' \emph{IEEE Transactions on
  Smart Grid}, vol.~11, no.~3, pp. 2464--2475, 2019.

\bibitem{8291006}
J.~{Silva}, J.~{Sumaili}, R.~J. {Bessa}, L.~{Seca}, M.~A. {Matos},
  V.~{Miranda}, M.~{Caujolle}, B.~{Goncer}, and M.~{Sebastian-Viana},
  ``Estimating the active and reactive power flexibility area at the tso-dso
  interface,'' \emph{IEEE Transactions on Power Systems}, vol.~33, no.~5, pp.
  4741--4750, 2018.

\bibitem{sun2019review}
H.~Sun, Q.~Guo, J.~Qi, V.~Ajjarapu, R.~Bravo, J.~Chow, Z.~Li, R.~Moghe,
  E.~Nasr-Azadani, U.~Tamrakar \emph{et~al.}, ``Review of challenges and
  research opportunities for voltage control in smart grids,'' \emph{IEEE
  Transactions on Power Systems}, vol.~34, no.~4, pp. 2790--2801, 2019.

\bibitem{9145713}
S.~{Karagiannopoulos}, C.~{Mylonas}, P.~{Aristidou}, and G.~{Hug}, ``Active
  distribution grids providing voltage support: The swiss case,'' \emph{IEEE
  Transactions on Smart Grid}, pp. 1--1, 2020.

\bibitem{8066335}
Z.~{Li}, Q.~{Guo}, H.~{Sun}, J.~{Wang}, Y.~{Xu}, and M.~{Fan}, ``A distributed
  transmission-distribution-coupled static voltage stability assessment method
  considering distributed generation,'' \emph{IEEE Transactions on Power
  Systems}, vol.~33, no.~3, pp. 2621--2632, 2018.

\bibitem{9165851}
E.~{Ucer}, M.~C. {Kisacikoglu}, and M.~{Yuksel}, ``Decentralized additive
  increase and multiplicative decrease-based electric vehicle charging,''
  \emph{IEEE Systems Journal}, pp. 1--9, 2020.

\bibitem{7164328}
L.~{Cheng}, Y.~{Chang}, and R.~{Huang}, ``Mitigating voltage problem in
  distribution system with distributed solar generation using electric
  vehicles,'' \emph{IEEE Transactions on Sustainable Energy}, vol.~6, no.~4,
  pp. 1475--1484, 2015.

\bibitem{7024906}
V.~{Aravinthan} and W.~{Jewell}, ``Controlled electric vehicle charging for
  mitigating impacts on distribution assets,'' \emph{IEEE Transactions on Smart
  Grid}, vol.~6, no.~2, pp. 999--1009, 2015.

\bibitem{shaukat2018survey}
N.~Shaukat, B.~Khan, S.~Ali, C.~Mehmood, J.~Khan, U.~Farid, M.~Majid, S.~Anwar,
  M.~Jawad, and Z.~Ullah, ``A survey on electric vehicle transportation within
  smart grid system,'' \emph{Renewable and Sustainable Energy Reviews},
  vol.~81, pp. 1329--1349, 2018.

\bibitem{shafiq2020reliability}
S.~Shafiq, U.~B. Irshad, M.~Al-Muhaini, S.~Z. Djokic, and U.~Akram,
  ``Reliability evaluation of composite power systems: Evaluating the impact of
  full and plug-in hybrid electric vehicles,'' \emph{IEEE Access}, vol.~8, pp.
  114\,305--114\,314, 2020.

\bibitem{8981892}
M.~R. {Islam}, H.~{Lu}, J.~{Hossain}, M.~R. {Islam}, and L.~{Li},
  ``Multiobjective optimization technique for mitigating unbalance and
  improving voltage considering higher penetration of electric vehicles and
  distributed generation,'' \emph{IEEE Systems Journal}, vol.~14, no.~3, pp.
  3676--3686, 2020.

\bibitem{yoon2017economic}
S.-G. Yoon and S.-G. Kang, ``Economic microgrid planning algorithm with
  electric vehicle charging demands,'' \emph{Energies}, vol.~10, no.~10, p.
  1487, 2017.

\bibitem{GONG2020665}
L.~Gong, W.~Cao, K.~Liu, Y.~Yu, and J.~Zhao, ``Demand responsive charging
  strategy of electric vehicles to mitigate the volatility of renewable energy
  sources,'' \emph{Renewable Energy}, vol. 156, pp. 665 -- 676, 2020.

\bibitem{8988201}
Z.~{Yang}, P.~{Dehghanian}, and M.~{Nazemi}, ``Seismic-resilient electric power
  distribution systems: Harnessing the mobility of power sources,'' \emph{IEEE
  Transactions on Industry Applications}, vol.~56, no.~3, pp. 2304--2313, 2020.

\bibitem{8386670}
S.~{Paul} and N.~P. {Padhy}, ``Resilient scheduling portfolio of residential
  devices and plug-in electric vehicle by minimizing conditional value at
  risk,'' \emph{IEEE Transactions on Industrial Informatics}, vol.~15, no.~3,
  pp. 1566--1578, 2019.

\bibitem{lu2018multi}
X.~Lu, K.~Zhou, S.~Yang, and H.~Liu, ``Multi-objective optimal load dispatch of
  microgrid with stochastic access of electric vehicles,'' \emph{Journal of
  cleaner production}, vol. 195, pp. 187--199, 2018.

\bibitem{li2020economic}
C.~Li, H.~Zhou, J.~Li, and Z.~Dong, ``Economic dispatching strategy of
  distributed energy storage for deferring substation expansion in the
  distribution network with distributed generation and electric vehicle,''
  \emph{Journal of Cleaner Production}, vol. 253, p. 119862, 2020.

\bibitem{humfrey2019dynamic}
H.~Humfrey, H.~Sun, and J.~Jiang, ``Dynamic charging of electric vehicles
  integrating renewable energy: a multi-objective optimisation problem,''
  \emph{IET Smart Grid}, vol.~2, no.~2, pp. 250--259, 2019.

\bibitem{lagarto2017optimisation}
M.~Lagarto, J.~F. Pinto, and L.~Ferreira, ``Optimisation of low-voltage
  distribution networks in a strong embedded microgeneration and electric
  vehicle penetration context,'' \emph{CIRED-Open Access Proceedings Journal},
  vol. 2017, no.~1, pp. 1922--1926, 2017.

\bibitem{de2014optimal}
J.~De~Hoog, T.~Alpcan, M.~Brazil, D.~A. Thomas, and I.~Mareels, ``Optimal
  charging of electric vehicles taking distribution network constraints into
  account,'' \emph{IEEE Transactions on Power Systems}, vol.~30, no.~1, pp.
  365--375, 2014.

\bibitem{aliasghari2020risk}
P.~Aliasghari, B.~Mohammadi-Ivatloo, and M.~Abapour, ``Risk-based scheduling
  strategy for electric vehicle aggregator using hybrid stochastic/igdt
  approach,'' \emph{Journal of Cleaner Production}, vol. 248, p. 119270, 2020.

\bibitem{9069461}
S.~{Nikkhah}, M.~A. {Nasr}, and A.~{Rabiee}, ``A stochastic voltage stability
  constrained ems for isolated microgrids in the presence of pevs using a
  coordinated uc-opf framework,'' \emph{IEEE Transactions on Industrial
  Electronics}, vol.~68, no.~5, pp. 4046--4055, 2021.

\bibitem{gengcoordinated}
L.~Geng, Z.~Lu, X.~Guo, J.~Zhang, X.~Li, and L.~He, ``Coordinated operation of
  coupled transportation and power distribution systems considering stochastic
  routing behaviour of electric vehicles and prediction error of travel
  demand,'' \emph{IET Generation, Transmission \& Distribution}, 2021.

\bibitem{ali2019voltage}
A.~Ali, D.~Raisz, and K.~Mahmoud, ``Voltage fluctuation smoothing in
  distribution systems with res considering degradation and charging plan of ev
  batteries,'' \emph{Electric Power Systems Research}, vol. 176, p. 105933,
  2019.

\bibitem{8332112}
IEEE, ``Ieee standard for interconnection and interoperability of distributed
  energy resources with associated electric power systems interfaces,''
  \emph{IEEE Std 1547-2018 (Revision of IEEE Std 1547-2003)}, pp. 1--138, 2018.

\bibitem{LI2021123625}
H.~Li, M.~Mao, K.~Guo, G.~Hao, and L.~Zhou, ``A decentralized optimization
  method based two-layer volt-var control strategy for the integrated system of
  centralized pv plant and external power grid,'' \emph{Journal of Cleaner
  Production}, vol. 278, p. 123625, 2021.

\bibitem{Soroudi2017}
A.~Soroudi, \emph{Power System Optimization Modeling in GAMS}.\hskip 1em plus
  0.5em minus 0.4em\relax Springer International Publishing, 2017.

\bibitem{zimmerman2010matpower}
R.~D. Zimmerman, C.~E. Murillo-S{\'a}nchez, and R.~J. Thomas, ``Matpower:
  Steady-state operations, planning, and analysis tools for power systems
  research and education,'' \emph{IEEE Transactions on power systems}, vol.~26,
  no.~1, pp. 12--19, 2010.

\bibitem{corliss2020electric}
T.~Dodson and S.~Slater, ``Electric vehicle charging behaviour study: final
  report for national grid eso,'' \emph{Element Energy Limited}, March, 2019.

\bibitem{pallonetto2020framework}
F.~Pallonetto, M.~Galvani, A.~Torti, and S.~Vantini, ``A framework for analysis
  and expansion of public charging infrastructure under fast penetration of
  electric vehicles,'' \emph{World Electric Vehicle Journal}, vol.~11, no.~1,
  p.~18, 2020.

\bibitem{G5}
Nationalgrid, \url{https://www.nationalgrid.com}, Online; accessed 1 July 2020.

\end{thebibliography}
\vfill\pagebreak
\end{document}